\documentclass[manuscript]{acmart} 

\usepackage{booktabs} 
\usepackage{color}
\usepackage{subfig}

\copyrightyear{2018} 
\acmYear{2018} 
\setcopyright{acmcopyright}
\acmConference[KDD '18]{The 24th ACM SIGKDD International Conference on Knowledge Discovery \& Data Mining}{August 19--23, 2018}{London, United Kingdom}
\acmBooktitle{KDD '18: The 24th ACM SIGKDD International Conference on Knowledge Discovery \& Data Mining, August 19--23, 2018, London, United Kingdom}
\acmPrice{15.00}
\acmDOI{10.1145/3219819.3219924}
\acmISBN{978-1-4503-5552-0/18/08}

\begin{document}
\title{Infrastructure Quality Assessment in Africa using Satellite Imagery and Deep Learning}

\author{Barak Oshri}
\affiliation{%
  \institution{Stanford University}
}
\email{boshri@stanford.edu}

\author{Annie Hu}
\affiliation{%
  \institution{Stanford University}
}
\email{anniehu@stanford.edu}

\author{Peter Adelson}
\affiliation{%
  \institution{Stanford university}
}
\email{padelson@stanford.edu}

\author{Xiao Chen}
\affiliation{%
  \institution{Stanford University}
}
\email{markcx@stanford.edu}

\author{Pascaline Dupas}
\affiliation{%
 \institution{Stanford University}
}
\email{pdupas@stanford.edu}

\author{Jeremy Weinstein}
\affiliation{%
  \institution{Stanford University}
}
\email{jweinst@stanford.edu}

\author{Marshall Burke}
%
\affiliation{%
  \institution{Stanford University}
}
\email{mburke@stanford.edu}

\author{David Lobell}
\affiliation{\institution{Stanford University}}
\email{dlobell@stanford.edu}

\author{Stefano Ermon}
\affiliation{\institution{Stanford University}}
\email{ermon@cs.stanford.edu}

\renewcommand{\shortauthors}{B. Oshri et al.}

\begin{abstract}
The UN Sustainable Development Goals allude to the importance of infrastructure quality in three of its seventeen goals. However, monitoring infrastructure quality in developing regions remains prohibitively expensive and impedes efforts to measure progress toward these goals. To this end, we investigate the use of widely available remote sensing data for the prediction of infrastructure quality in Africa. We train a convolutional neural network to predict ground truth labels from the Afrobarometer Round 6 survey using Landsat 8 and Sentinel 1 satellite imagery. 

Our best models predict infrastructure quality with AUROC scores of 0.881 on Electricity, 0.862 on Sewerage, 0.739 on Piped Water, and 0.786 on Roads using Landsat 8. These performances are significantly better than models that leverage OpenStreetMap or nighttime light intensity on the same tasks. We also demonstrate that our trained model can accurately make predictions in an unseen country after fine-tuning on a small sample of images. Furthermore, the model can be deployed in regions with limited samples to predict infrastructure outcomes with higher performance than nearest neighbor spatial interpolation.

\end{abstract}



\maketitle

\section{Introduction}
Basic infrastructure availability in developing regions is a crucial indicator of quality of life \cite{pottas2014addressing}. Reliable infrastructure measurements create opportunities for effective planning and distribution of resources, as well as guiding policy decisions on the basis of improving the returns of infrastructure investments \cite{varshney2015targeting}.
Currently, the most reliable infrastructure data in the developing world comes from field surveys, and these surveys are expensive and logistically challenging \cite{dabalen2016mobile}. Some countries have not taken a census in decades \cite{xie2016transfer}, and data on key measures of infrastructure development 
are still lacking for much of the developing world \cite{jean2016combining, ieag2014world}.  Overcoming this data deficit with more frequent surveys is likely to be both prohibitively costly, perhaps costing hundreds of billions of U.S. dollars to measure every target of the United Nations Sustainable Development Goals in every country over a 15-year period \cite{jerven2014benefits}, and institutionally difficult, as some governments see little benefit in having their performance documented \cite{sandefur2015political,jean2016combining}.

\begin{figure*}[!hbpt]
\centering
\hspace*{-0.4cm}
\begin{tabular}{cccc}
\raisebox{5.5\normalbaselineskip}[0pt][0pt]{\rotatebox{90}{Labels}} & 
{\includegraphics[scale=0.35]{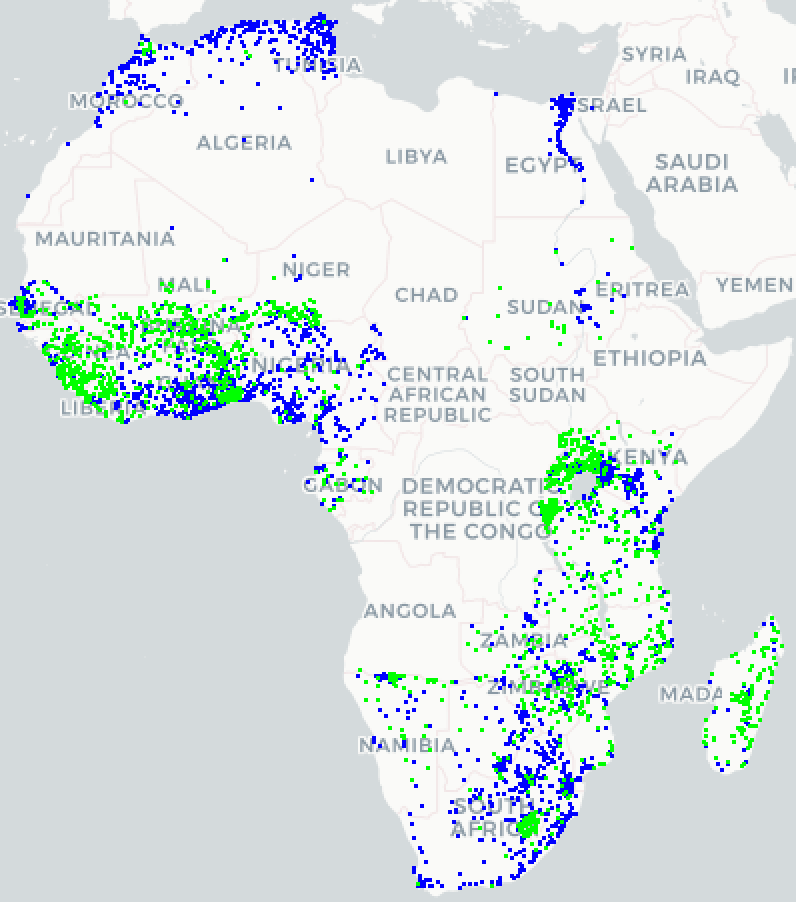}} &
{\includegraphics[scale=0.35]{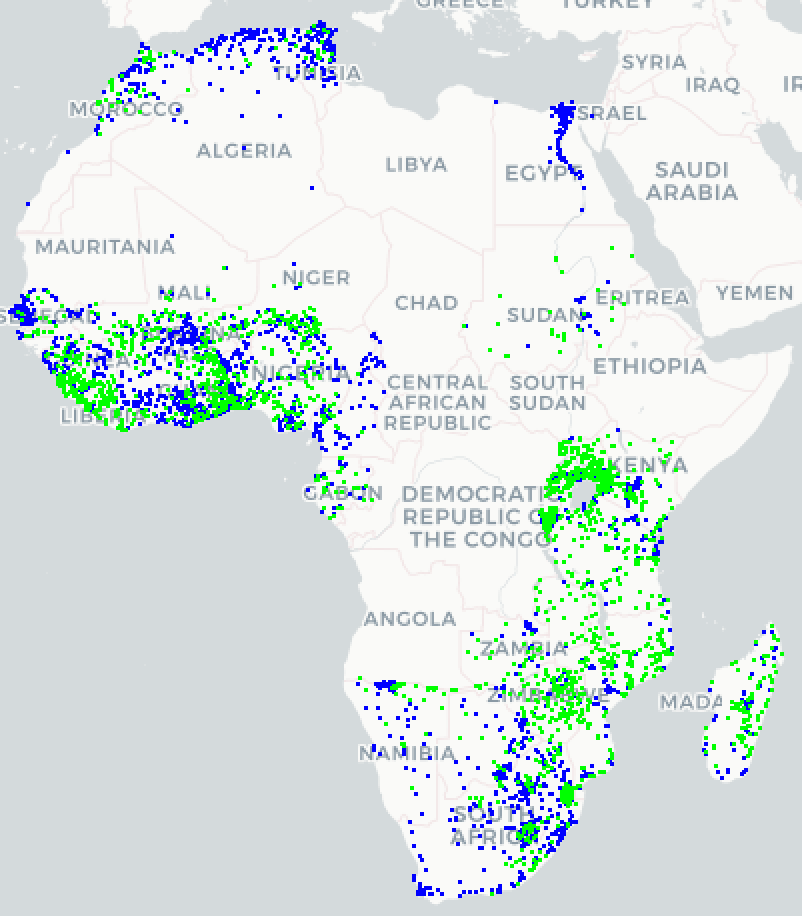}} &
{\includegraphics[scale=0.35]{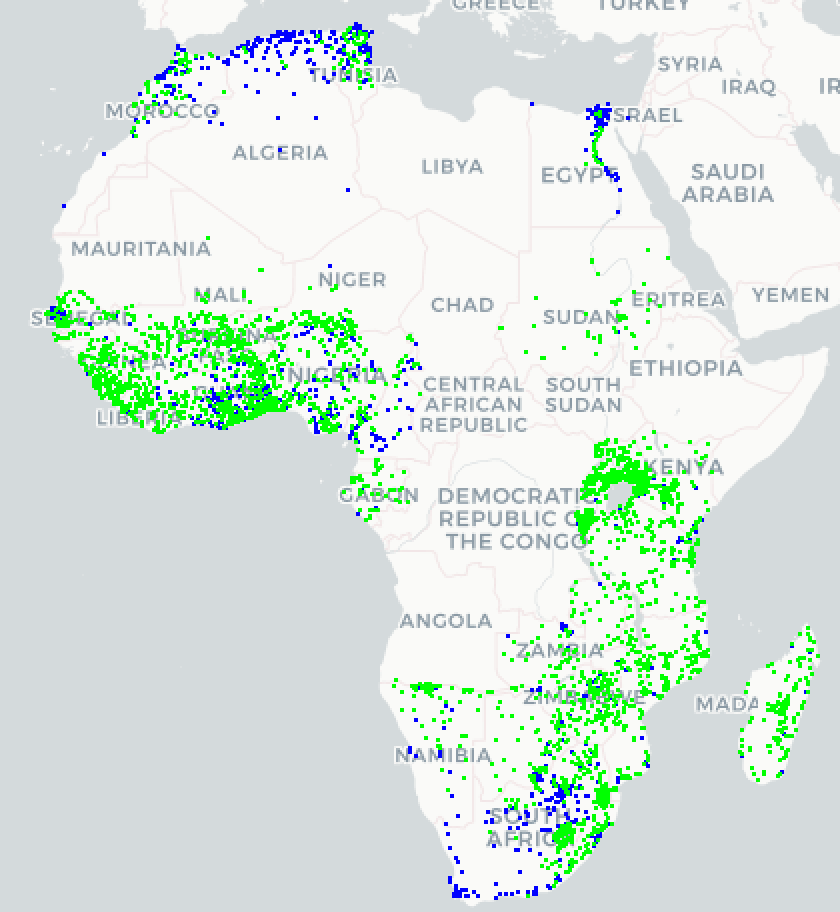}} \\
\raisebox{5.1\normalbaselineskip}[0pt][0pt]{\rotatebox{90}{Predictions}} & 
{\includegraphics[scale=0.35]{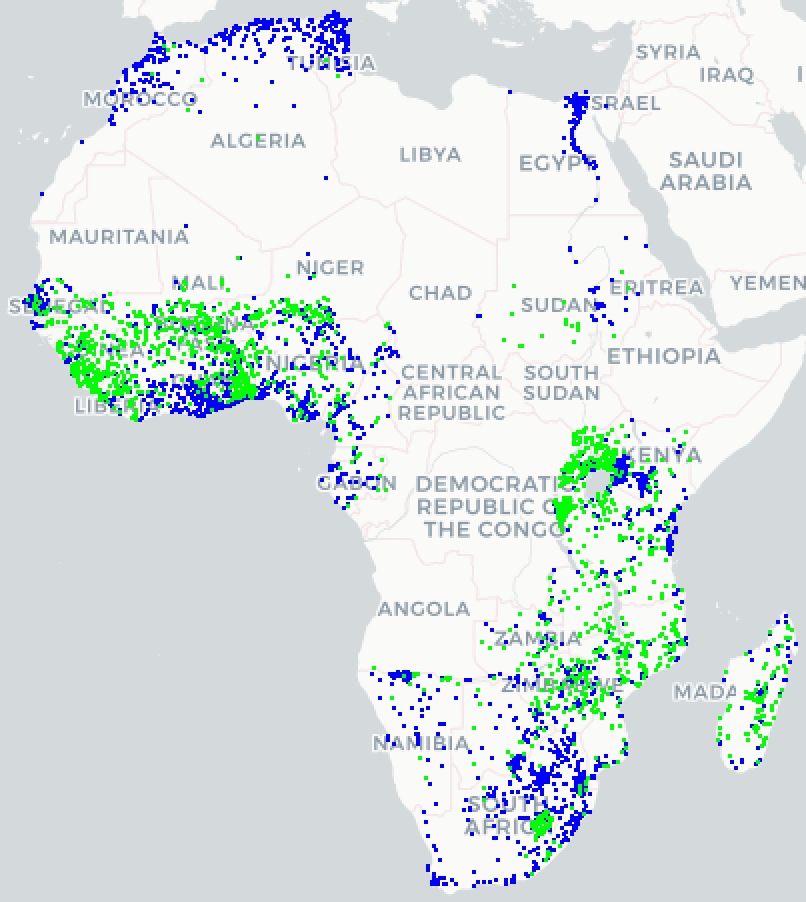}} &
{\includegraphics[scale=0.35]{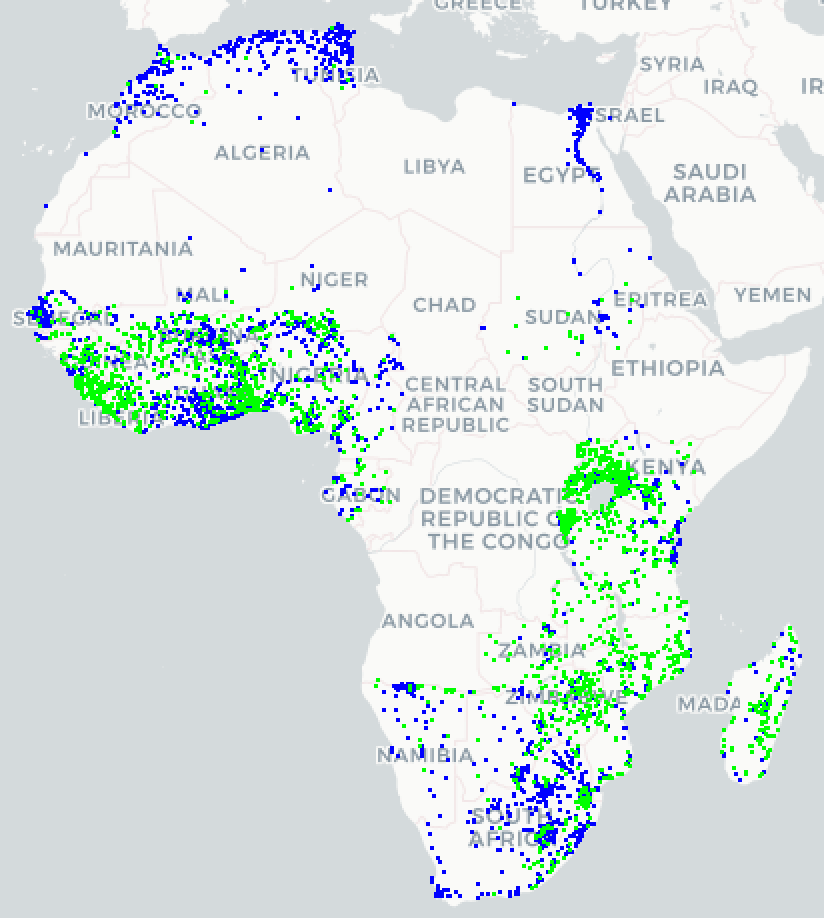}} &
{\includegraphics[scale=0.35]{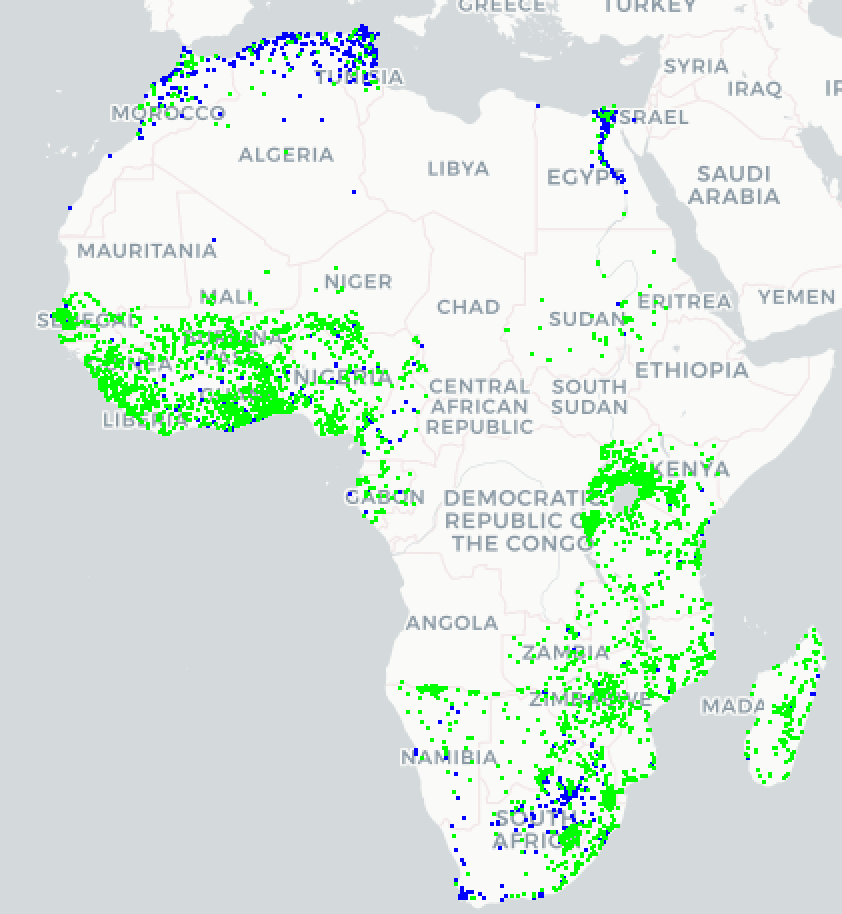}} \\
& Sewerage & Electricity & Piped Water
\end{tabular}
\caption{We compare the labels for \emph{sewerage}, \emph{electricity}, and \emph{piped water}, in the top row, with our predictions for these variables in the bottom row. Positive labels are shown in blue and negative labels in green.}
\label{best_vis}
\end{figure*}

One emerging technology for the global observation of infrastructure quality is satellite imagery. As satellite monitoring becomes more ubiquitous, with an increasing number of commercial players in the sector, improvements in spatial and temporal resolution open up new applications, uses, and markets \cite{/content/publication/9789264217294-en}, including the possibility to monitor important sustainability outcomes at scale. Additionally, satellite imagery can observe developing countries that do not have significant prior data, containing a wealth of observations that can be harnessed for social development applications \cite{jean2016combining}, including infrastructure assessment.

Such rich and high quality image data enable advanced machine learning techniques to perform sophisticated tasks like object detection and classification, and deep learning in particular has shown great promise \cite{esteva2017dermatologist, he2016deep, dai2016r, albert2017using}. While a number of recent papers discuss the use of deep learning on satellite imagery for applications in land use cover \cite{albert2017using}, urban planning \cite{audebert_beyond_2017}, environmental science \cite{bragilevskydeep}, etc. \cite{DBLP:journals/corr/abs-1710-05483,you2017deep,pryzant2017monitoring}, many unanswered questions remain in the field, particularly in the application of deep learning to social and economic development. 

Our contributions in this paper are to both the applied deep learning literature and to socioeconomic studies involving remote sensing data. We propose a new approach to using satellite imagery combined with field data to map infrastructure quality in Africa at the 10m and 30m resolution. We explore multiple infrastructure outcomes, including but not limited to \emph{electricity}, \emph{sewerage}, \emph{piped water}, and \emph{road} to identify the remote sensing predictability of different infrastructure categories on a continental level. Prediction maps for three outcomes are given in Figure 1. We show that, through fine tuning a pretrained convolutional neural network (CNN), our models achieve 0.881, 0.862, 0.739, and 0.786 area of the receiver operating characteristic (AUROC) scores on these outcomes and perform better than nighttime lights intensity (nightlights) and OpenStreetMap (OSM). Our primary datasets are a combination of 10m and 30m resolution satellite imagery from Sentinel 1 and Landsat 8 respectively as well as the georeferenced Afrobarometer Round 6 survey encompassing 36 countries in Africa. Our work provides the ability to assess infrastructure in an accurate and automated manner, to supplement the spatial extent of field survey data, and to generate predictions in unseen regions.

To the best of our knowledge, we are the first to use CNNs with Sentinel 1 imagery for social development research.

\subsection{Organization of Paper}
The remainder of the paper is organized as follows. Section 2 (Related Work) discusses recent applications of machine learning on satellite imagery and contextualizes previous work in infrastructure quality detection. Section 3 (Data) describes the survey data and satellite imagery data sources. Section 4 (Methodology) introduces the problem formulation and modeling techniques used in this paper. Section 5 (Experimental Results) presents the performance of our model. Section 6 (Baseline Models) benchmarks our model performance against three baselines. In Section 7 (Generalization Capabilities) we explore a few settings that test the deployment potential of the model, including its performance on urban and rural enumeration areas, as well as performance in countries that the model was not originally trained on. In Section 8 we discuss conclusions and future work.

\section{Related Work}

\noindent The application of CNNs to land use classification can be traced back to the work of \citet{Castelluccio2015land} and \citet{penatti2015deep} who trained deep models on the UC Merced land use dataset \cite{yang2010bag}, which consists of 2100 images spanning 21 classes. Similar early studies on land use classification that employ deep learning techniques are the works of \citet{romero2016unsupervised} and \citet{papadomanolaki2016benchmarking}. In \citet{liu2017learning} a spatial pyramid pooling technique is employed for land use classification using satellite imagery. These studies adapted architectures pre-trained to recognize natural images from the ImageNet dataset, such as VGGNet \cite{simonyan2014very}, to fine-tune them on their much smaller land use data. A more recent study \cite{albert2017using} uses state-of-the-art deep CNNs VGG-16 \cite{simonyan2014very} and Residual Neural Networks \cite{he2016deep} to analyze land use in urban neighborhoods with large scale satellite data. 

A few recent works, which are related to infrastructure detection through deep learning, inspire us to use additional data sources such as OSM \cite{haklay2008openstreetmap} to support our investigation. One project that is closely related to our investigation is DeepOSM\footnote{https://github.com/trailbehind/DeepOSM}, in which the authors take the approach of pairing OSM labels with satellite imagery obtained from Google Maps and use a convolutional architecture for classification. In \cite{yuan2016automatic}, the authors show that their model can achieve a precision of 0.74 and recall of 0.70 on building detection, training CNNs on 0.3 meter resolution OSM images. Their CNN consisted of 7 identical blocks of of filtering, pooling, and convolutional layers. \citet{mnih2010learning, mnih2012learning} built satellite image models for road detection, and they obtain almost 0.8 precision and 0.9 at best case in one urban area. 
Recently, \cite{albert2017using} predicted land use classes in urban environments with 0.7 to 0.8 accuracy, commenting on the inherent difficulty in the task of understanding high-level, subjective concepts of urban planning from satellite imagery. 

Compared to prior work, our key contributions are that we train the first wide scale classifier of infrastructure quality via deep learning and publicly available satellite imagery on 11 infrastructure quality outcomes, and that we are able to achieve state-of-the-art performance in predicting infrastructure accessibility on a large imagery dataset in Africa. 


\section{Data}
This project relates two data sources in a supervised learning setting: survey data from Afrobarometer as ground truth infrastructure quality labels, and satellite imagery from Landsat 8 and Sentinel 1 as input sources.
%
%
\subsection{Afrobarometer Round 6}
\begin{figure}[h!]
\centering
\includegraphics[scale=0.85]{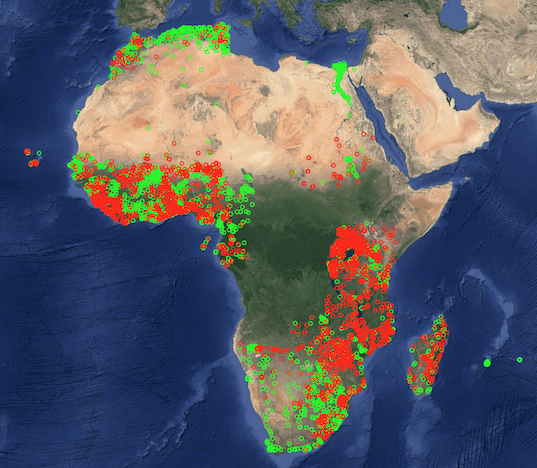}
\caption{Distribution of \emph{piped water} in the Afrobarometer Round 6 survey. Our study is the first to conduct a large scale infrastructure analysis at the continent-scale across 36 nations in Africa. Positive examples are shown in green and negative examples in 0.}
\label{cs235b:report:afro:vis}
\end{figure}
\noindent Conducted over 2014 and 2015 across 36 countries in Africa, the Afrobarometer Round 6 survey collected surveyor-assessed quality indicators about infrastructure availability, access, and satisfaction based on respondent data \footnote{Available by request with AidData} \cite{benyishay2017geocoding}. The dataset surveys 7022 enumeration areas with 36 attributes regarding various aspects of welfare, infrastructure quality, and wealth. \cite{AfrobarometerElectricityReport}. Afrobarometer data from previous rounds dates back to 1999; our application used only Round 6, but adding other rounds represents valuable further work.  Each survey response is an aggregate of face-to-face surveys in the enumerated area, which can encompass a city, village, or rural town, and between 1200 and 2400 samples are collected over all enumeration areas for each country. Each country in the Round 6 survey has between 150 and 300 enumeration areas. Each enumeration area is georeferenced with its latitude and longitude, and we center a satellite image for each enumeration area around these coordinates \cite{AfrobarometerManual}. Figure \ref{cs235b:report:afro:vis} shows the spatial distribution of all enumeration areas.

The Afrobarometer Round 6 survey includes 11 binary infrastructure outcomes, with each denoting the availability and quality of that infrastructure in the enumeration area. We primarily focus on highlighting results in \emph{electricity}, \emph{sewerage}, \emph{piped water}, and \emph{road} for their novel contributions. We show results on all binary outcomes except for \emph{cellphone} and \emph{school} due to their high class imbalances. Due to the variation of class balances across all variables we assess performance on multiple metrics and stress AUROC due to its insensitivity to class imbalances. This helps assess comparability in performance between the outcomes.

\begin{table}[h!]
\centering
\caption{Overview of number of examples in each class for all binary infrastructure variables in the Afrobarometer Round 6 survey, including the balance (proportion of positive labels).}
\label{afro_number}
\begin{tabular}{|c|c|c|c|}
\hline
Infrastructure & Label 1 & Label 0 & Balance  \\
\hline
Electricity & 4680& 2343 & 0.667  \\
Sewerage &2239& 4784 & 0.319 \\
Piped Water &  4303& 2720 & 0.613 \\
Road &  3886& 3137 & 0.553 \\
Post Office&  1728& 5295 & 0.246 \\
Market Stalls&  4811& 5148 & 0.685 \\
Police Station&  2553& 4470 & 0.364\\
Bank&  1875& 0.767 & 0.267 \\
Cellphone & 6576 & 456 & 0.936\\
School & 6082 & 941 & 0.866\\
Health Clinic & 4115 & 2908 & 0.586 \\
\hline 
\end{tabular}
\end{table}

\subsection{Satellite Imagery}
\noindent Two primary sources of satellite observations were used, both offering coverage of most of the enumeration areas. The satellite data is temporally consistent with the survey data, from 2014 and 2015. For a given enumeration area with sampling location (latitude, longitude) at the center, we collect $500\times 500$ pixel images.
\\\\
\noindent \textbf{Landsat 8}: Landsat 8 is a satellite with the objective of collecting publicly available multispectral imagery of the global landmass. Landsat 8 imagery has a 30m resolution providing $15km \times 15km$ coverage in the following six bands:  Blue, Green, Red, Near Infrared, and two bands of Shortwave Infrared. Each pixel value represents the degree of reflectance in that specific band. Cloud cover removal is handled natively by the Landsat 7 Automatic Cloud Cover Assessment algorithm \cite{irish2000landsat}.
\\\\
\noindent \textbf{Sentinel 1}: Sentinel 1 is the first satellite in the Copernicus Programme satellite constellation. It uses a C-band synthetic-aperture radar (SAR) instrument to acquire imagery regardless of weather and light conditions. Imagery obtained from Sentinel 1 has a resolution of 10 meters, providing $5km \times 5km$ coverage. It is processed to five bands, comprised of four polarizations and a look angle: VV, VH, Angle, VV$_{\gamma^0}$, and VH$_{\gamma^0}$. Each pixel value in the polarization channels represents the degree of backscatter in that specific band. For the Afrobarometer dataset, images were taken from two different orbital paths resulting in different look angles, ascending or descending, though not every enumerated area had both images. We choose the image with ascending path when available, otherwise we choose the image with descending path.

\section{Methodology}

\subsection{Problem Formulation}
The infrastructure detection task is a multi-label binary classification problem. The input is a satellite image $X$ and the outputs are binary labels $Y_1, ..., Y_k \in \{0, 1\}$, corresponding to quality indicators of different infrastructure outcomes. We optimize the mean binary cross entropy loss
\begin{equation}
L(X, {Y_1, ..., Y_k}) = \frac{1}{k}\sum_{i=1}^k Y_i\log p(Y_i = 1 | X) + (1 - Y_i)\log p(Y_i = 0 | X)
\end{equation}
\noindent where $p(Y_i = j | X)$ is the probability that the model predicts that input $X$ has label $j$ for infrastructure outcome $Y_i$. 

\subsection{Model}
We train deep learning models to learn useful representations of data from input imagery. Convolutional Neural Networks (CNNs) have been particularly successful in vision tasks and have also been demonstrated to perform well on satellite imagery \cite{xie2016transfer, jean2016combining, albert2017using, bragilevskydeep}. For all experiments in the paper, we train a Residual Neural Network (ResNet) architecture \cite{he2016deep}. The following describe further specifications of our model:

\textbf{ResNet}. ResNet has achieved state-of-the-art results in ImageNet \cite{he2016deep}, and its main contribution over previous convolutional neural networks is learning residual functions in every forward propagation step with reference to the layer inputs. We posit that this is useful in satellite imagery analysis for retaining low-level features in high-level classifications. We train an 18 layer network.

\textbf{Transfer Learning}. Instead of training the network from random initializations, we initialize our network weights with those of a ResNet pre-trained on ImageNet \cite{krizhevsky2012imagenet}. Even though the weights are initialized on an object recognition task, this approach has been demonstrated to be effective in training on new tasks compared with initializing using random weights \cite{oquab2014learning} and useful for learning low-level features like edges in satellite imagery.

\textbf{Multi Channel Inputs}. ImageNet architectures originally take inputs with three channels corresponding to RGB values. However, Landsat 8 and Sentinel 1 have six and five bands respectively. We change the first convolution layer in the network to have greater than three input channels by extending the convolutional filters to further channels. The number of output channels, stride, and padding for the first layer is the same as in the original ResNet. With Landsat 8, we initialize the RGB band parameters of the first layer layer with the same parameters as in the pre-trained ResNet weights, and initialize the non-RGB bands with Xavier initialization \cite{glorot2010understanding}. With Sentinel 1, which does not include RGB bands, we initialized three bands with the pretrained RGB channel weights, and the other two with Xavier initialization. In Xavier initialization, each weight $W$ is sampled uniformly from 
\begin{equation}
W \sim U\bigg[\frac{\sqrt{6}}{n_{\text{in}} + n_{\text{out}}}, \frac{\sqrt{6}}{n_{\text{in}} + n_{\text{out}}}\bigg]
\end{equation}

\noindent where $n_{\text{in}}$ is the input dimensions of the convolutional filter and $n_{\text{out}}$ is the number of neurons in the next layer.

\subsection{Data Processing}
Our pipeline includes several data processing steps and augmentation strategies:

\textbf{Unique Geocoded Images in Test Set}. Each enumerated area in the Afrobarometer survey has a unique geocode field, and enumerated areas with the same geocode field have substantial spatial overlap. To ensure that there is no spatial overlap between images observed in the training set and in the test set, we enforce that only points with a unique geocode appear in the test set.

\textbf{Cropping}. Our satellite images are ingested at $500 \times 500$ pixel bounding boxes. We try downsampling, cropping at random regions, and cropping around the center pixel to $224 \times 224$ pixels, and we find that the latter has best performance and convergence.

\textbf{Horizontal Flipping}. To augment the limited size of our dataset, at training time we rotate the image around the x-axis with 50\% probability. 

\textbf{Normalization}. We normalize our data channel-wise to zero mean and unit standard deviation.

\begin{table*}[!htb]
\centering
\caption{Our Landsat 8 model achieves AUROC scores above 0.85 on \emph{electricity} and \emph{sewerage}, and achieves scores above 0.7 on all but three outcome variables.}
\label{afro_test}
\begin{tabular}{|c|c|c||c|c|c|c|c|c|}
\hline
Satellite & Infrastructure & Balance & Accuracy & F1 Score & Precision & Recall & AUROC  \\
\hline
L8 & \textbf{Electricity} & 0.667& 0.832 & 0.873 & 0.877 & 0.870  & \textbf{0.881} \\
L8 & \textbf{Sewerage} &0.319& 0.815 & 0.700& 0.756 & 0.650  & \textbf{0.862} \\
L8 & \textbf{Piped Water}&  0.613& 0.673 & 0.725& 0.730 & 0.720  & \textbf{0.739} \\
L8 & \textbf{Road}&  0.553& 0.705 & 0.704& 0.746 & 0.667  & \textbf{0.786} \\
L8 &Bank&  0.267& 0.767 & 0.364& 0.543 & 0.273 & 0.726 \\
L8 & Post Office&  0.246& 0.753 & 0.427& 0.434 & 0.420  & 0.712 \\
L8 &Market Stalls&  0.685& 0.681 & 0.791& 0.688 &0.930  & 0.665 \\
L8 &Health Clinic&  0.586& 0.622 & 0.719& 0.632 & 0.833 & 0.664 \\
L8 &Police Station&  0.364& 0.660 & 0.492& 0.490& 0.494  & 0.650\\
\hline
S1 & Electricity & 0.667& 0.769& 0.820& 0.820& 0.821  & 0.819 \\
S1 & \textbf{Sewerage} &0.319& 0.802 & 0.659& 0.678 & 0.842  & \textbf{0.862} \\
S1 & Piped Water& 0.613 & 0.663 & 0.722& 0.716 & 0.728  & 0.725\\
S1 & Road& 0.553 & 0.702 & 0.730& 0.681 & 0.786  & 0.779\\
\hline 
\end{tabular}
\end{table*}

\subsection{Training}
We train independent models for Landsat 8 and Sentinel 1. Our models are trained with multi-label classifiers as well as single-label classifiers, and we report the higher performance for each variable. We train the network end-to-end using Adam optimizer with $\beta_1 = 0.9$ and $\beta_2 = 0.999$ \cite{kingma2014adam}. We train the model with a batch size of 128, and we update our weights with a decaying learning rate of 0.0001. The model weights are regularized with L2 regularization of 0.001.

Due to the limited size of our dataset, we evaluate our model on all the data with K-fold cross validation where $K=5$, producing a train-test split for every fold with 80\% training and 20\% testing. We train a model for each fold, predict values on the test set, and once every fold has been tested compute our evaluation metrics. 

\section{Experimental results}

\subsection{Evaluation Metrics}
Performance of the model was evaluated by a number of metrics:  accuracy, F1-score, precision, recall, and AUROC \cite{ROC}. F1 is calculated as the harmonic mean of precision and recall.  AUROC corresponds to the probability that a classifier will rank a randomly chosen positive example higher than a randomly chosen negative example and generally ranges between 0.5 being a random classifier and 1.0 being a perfect predictor. 

\subsection{Classification Results}

In Table \ref{afro_test} we show the classification results our model achieves on each category. Landsat 8 performs better than Sentinel 1 on every category; we show results for Sentinel 1 on our four highest scoring categories. The first column displays the proportion of images that have label 1 in that category to compare with the accuracy our model achieves. 

On \emph{electricity} and \emph{sewerage}, our best model achieves AUROC greater than 0.85. In particular, the model achieves an F1 score of 0.873 on \emph{electricity}. Our model does not perform as effectively on other variables, such as \emph{market stalls}, \emph{health clinic}, and \emph{police station}. This is not surprising since Landsat 8 and Sentinel 1 operate at a resolution lower than that needed to resolve individual objects that signify the presence of these facilities. With the better performing categories, the imagery still cannot resolve individual electricity lines, roads, or water tanks at 30m resolution; however, the structures in aggregate might contribute to different spectral signatures. This means that the classification is likely relying on large-scale proxies, such as urban sprawl and geographical features, that correlate with the class values.

\section{Baseline Models}

We compare our model performance with several baselines of different input sources. To evaluate the difficulty of the task, we also compare against an ideal baseline that uses (expensive to collect) survey labels to make predictions. We suggest that the oracle defines a reasonable notion of great performance on this dataset. 


\begin{table}[h!]
\centering
\caption{We compare our model with four baselines on AUROC scores. Our Landsat 8 models outperform nightlights and OSM models and performs slightly better or comparably with nearest neighbor spatial interpolation. Performance on three infrastructure outcomes is comparable with the oracle.}
\label{baseline_results}
\begin{tabular}{c|ccc|c|c}
\hline 
Infrastructure & Nightlights & OSM & Spatial & Oracle & L8  \\
\hline
\hline 
Electricity & 0.79 & 0.73 & 0.78 & 0.89 & \textbf{0.88} \\
Sewerage & 0.75 & 0.77 & 0.78 & 0.89 & \textbf{0.86} \\
Piped Water & 0.73 & 0.73 & 0.75 & 0.89 & \textbf{0.74} \\
Road & 0.67 & 0.68 & 0.74 & 0.79 & \textbf{0.79} \\
Bank & 0.57 & 0.70 & 0.67 & 0.93 & \textbf{0.73} \\
Post Office & 0.56 & 0.64 & 0.70 & 0.92 & \textbf{0.71} \\
Market Stalls & 0.50 & 0.62 & 0.66 & 0.84 & \textbf{0.66} \\
Health Clinic & 0.52 & 0.61 & 0.64 & 0.85 & \textbf{0.66} \\
Police Station & 0.54 & 0.63 & 0.66 & 0.90 & \textbf{0.65} \\
\hline
\end{tabular}
\end{table}
\begin{figure*}[!htb]
\centering
  \begin{minipage}{0.22\textwidth}
    \centering
    \includegraphics[scale=0.23]{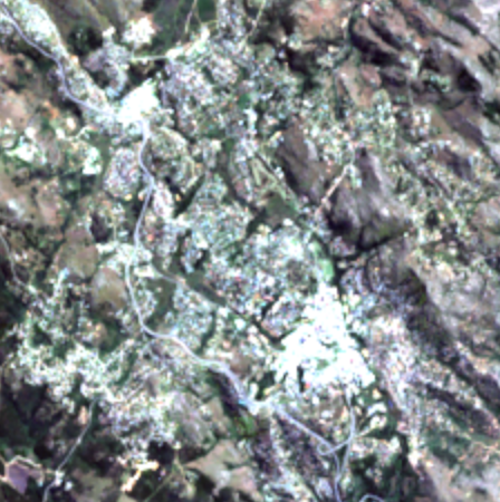}
  \end{minipage}
  \begin{minipage}{0.22\textwidth}
  	\centering
    \includegraphics[scale=0.23]{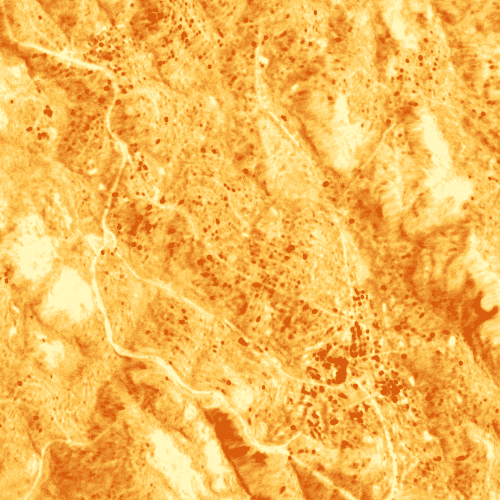}
  \end{minipage}
   \begin{minipage}{0.22\textwidth}
  	\centering
    \includegraphics[scale=0.23]{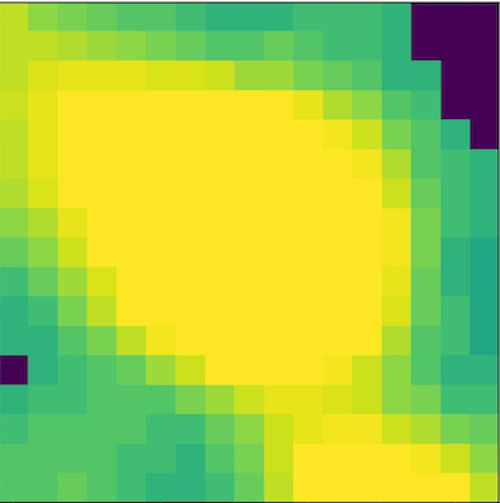}
  \end{minipage}
    \begin{minipage}{0.22\textwidth}
  	\centering
    \includegraphics[scale=0.23]{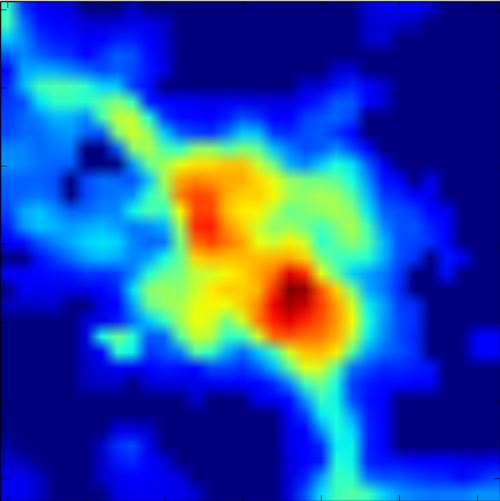}
  \end{minipage}
   \caption{Visualization of four satellite imagery sources we use in this paper. Each image is centered around the same geolocation. From left to right: Landsat 8 (multispectral), Sentinel 1 (synthetic aperture radar), DMSP (coarse nightlights), and VIIRS (nightlights). We find that when coupled with the neural network architectures we considered, Landsat 8 is the most informative source of information about infrastructure quality, followed by Sentinel 1.}
  \label{satellites}
\end{figure*}
\subsection{Nightlights Intensity}
\citet{jean2016combining} used nighttime light intensities as a proxy for poverty level. Since poverty and infrastructure are closely related, we use nighttime lights as a baseline predictor for infrastructure level. For example, we expect nightlight intensity to be a good proxy for electricity access.
We use nighttime light intensity data from the Defense Meteorological Satellite Program (DMSP) \cite{dmsp}, imaged in 2013 with a resolution of 30 arc-seconds, and Visible Infrared Imaging Radiometer Suite (VIIRS) \cite{viirs}, imaged in 2015 with a resolution of 15 arc-seconds.

For each survey response, we take a $7\times 7$ DMSP or $14\times 14$ VIIRS patch of pixels centered on the geolocation. For both sources, this corresponds to roughly a $6km$ square, which matches the area coverage of the cropped Landsat images we used in our best model. Figure~\ref{satellites} visualizes all four satellite imagery sources used in this paper for one geolocation. We run a logistic regression classifier for each response variable using cross-validated parameters, and we take the prediction of the better-performing nightlights satellite for each variable. Table~\ref{baseline_results} shows full results from this baseline. 

As expected, nightlights perform quite well at predicting \emph{electricity} (AUROC of 0.79), and has some predictive power with water, sewerage and roads. However, its performance in other outcomes is only slightly better than random chance (AUROC only a little better than 0.5).  Using nightlights thus offers only a limited window into infrastructural provisions by proxying human activity as light emissions and fails to attend to facilities that may be present without such evidence.

\subsection{OpenStreetMap}
OpenStreetMap (OSM) is a collaborative project for creating a map of the world with crowdsourced information. Users and organizations upload georeferenced tags about anything they would like to identify on the map. OSM contains a wealth of information on infrastructure where it is available. However, because of its crowdsourced nature, the data is less reliable compared to professional surveys \cite{helbich2012comparative}. 

For each enumeration area in a $15km \times 15km$ bounding box around the center geocoordinate, we extract the total number of highways and buildings in OSM. We expand the set of input features for every area with several non-linear transformations on these counts, including $log$, \emph{square root}, and highway-to-building $ratios$. We normalize each feature to zero mean and unit standard deviation, and then train logistic regression, support vector machine, and random forest classifiers on the set of input features. The logistic regression performs best.


We display results in Table~\ref{baseline_results}.  
With AUROCs above 0.7 for \emph{electricity}, \emph{piped water} and \emph{sewerage}, results indicate meaningful predictive power. OSM performs worse than nightlights in electricity access, but is generally better for other tasks. Surprisingly, the OSM features are not predictive of the \emph{road} category (AUROC 0.68). Overall, our results indicate that although OSM is imperfect, it does provide some useful insight into infrastructure quality metrics, achieving AUROCs between $0.62$ and $0.77$, indicating good performance at discriminating high vs. low infrastructure quality enumeration areas across the African continent. 

\subsection{Spatial Interpolation}
We also compute the baseline performance on the infrastructure outcomes using spatial interpolation methods. For each enumeration area we consider how predictive its latitude and longitude are of each infrastructure variable nonparametrically. We uniformly sample 80\% of the enumeration areas as the training set, and for each survey response in the test set use nearest neighbor grid interpolation, which labels the example with the label of its closest neighbor. We take the infrastructure value of the neighbor as our predicted value. 

Our spatial interpolation model achieves better performance than Nightlights and OSM models but has lower performance compared with the Landsat 8 models on the highest performing infrastructure outcomes, especially in \emph{electricity} and \emph{sewarage}. This indicates that though geographic location is a non-negligible predictor of infrastructure development, satellite imagery is able to extract deeper and more useful insights. Additionally, since spatial interpolation methods are one long-established approach in survey data interpolation \cite{reibel2007geographic}, we suggest that our model can be used to improve how survey samples are interpolated to larger regions. 

\begin{figure*}[!htb]
\centering
  \begin{minipage}{0.22\textwidth}
    \centering
    \includegraphics[scale=0.22]{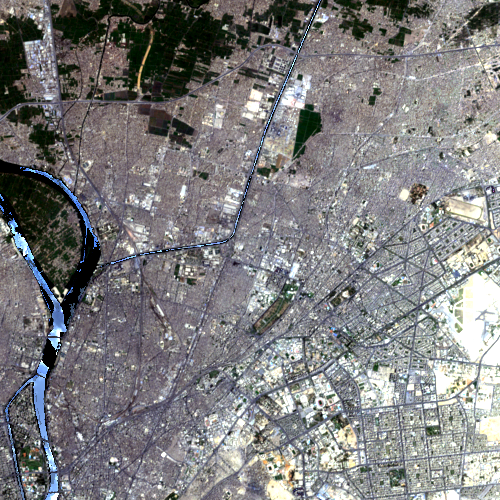}
  \end{minipage}
  \begin{minipage}{0.22\textwidth}
  	\centering
    \includegraphics[scale=0.22]{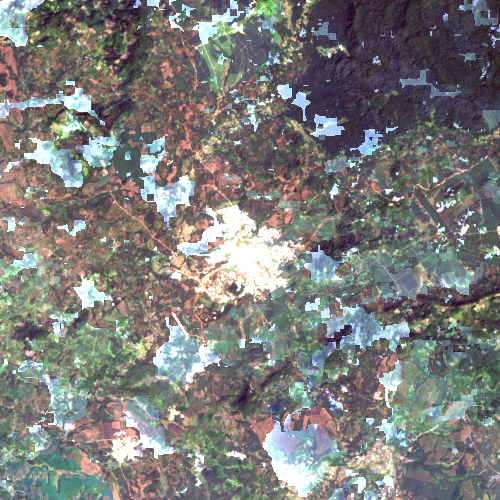}
  \end{minipage}
   \begin{minipage}{0.22\textwidth}
  	\centering
    \includegraphics[scale=0.295]{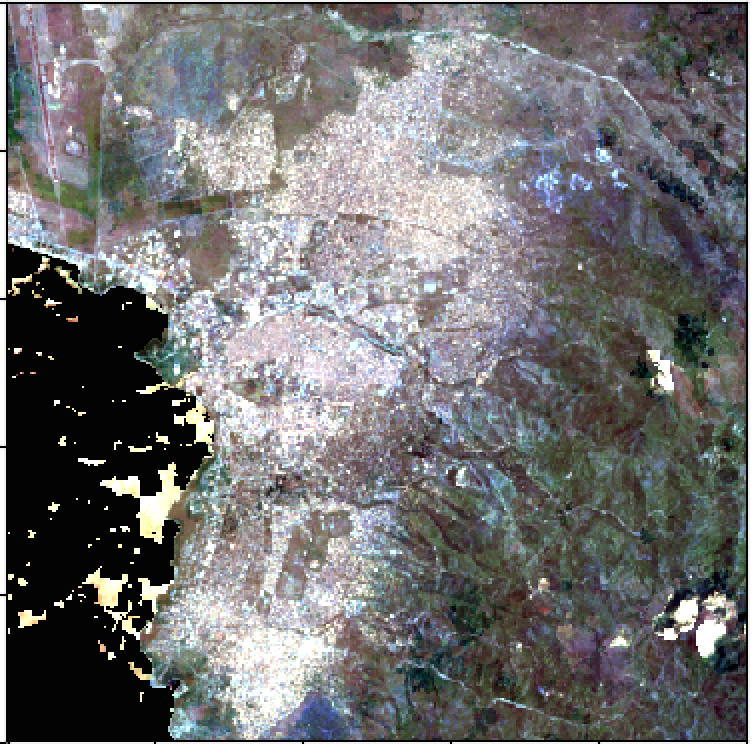}
  \end{minipage}
    \begin{minipage}{0.22\textwidth}
  	\centering
    \includegraphics[scale=0.297]{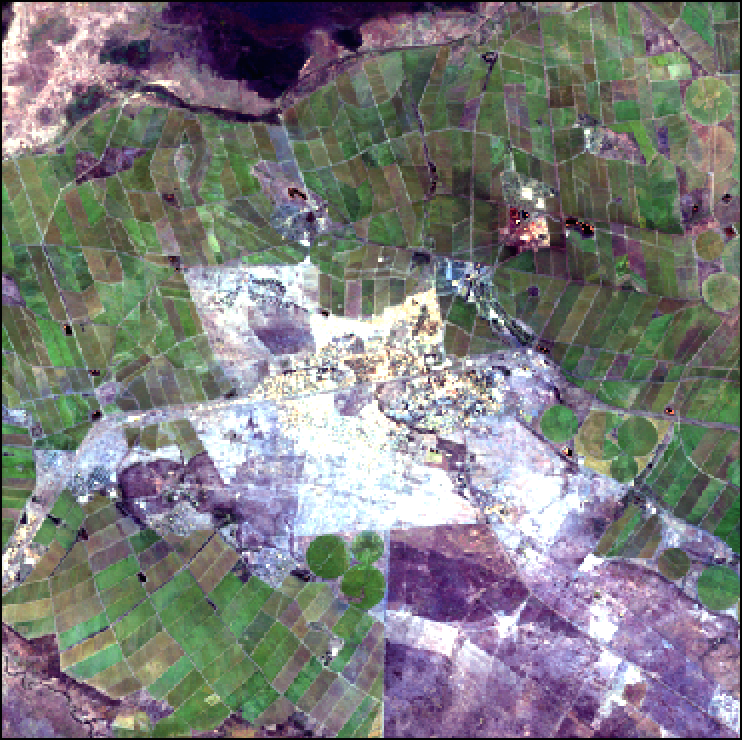}
  \end{minipage}
\caption{Predictions from left to right: true positive \emph{piped water} (Egypt, urban), false positive \emph{piped water} (Malawi, rural), true positive \emph{electricity} (Burundi, urban), false positive \emph{electricity}. (Burkino Faso, rural) }
\label{vis}
\end{figure*}

\subsection{Cross-label Predictions}
Finally, we construct an \textit{oracle} baseline to assess the discriminatory performance between the high quality infrastructure labels. In this baseline, we use all observed variables in the survey data to predict one held-out variable. That is, if there are $n$ infrastructure quality variables, we learn parameters $W_{ij}$ such that for all $i, j \leq n$, 
\begin{equation}
\tilde{Y_i} = \sigma\bigg(\sum_{j=1}^n W_{ij}Y_j\bigg)
\end{equation}
where $\sigma$ is the sigmoid function. We optimize $Y_i$ and $\tilde{Y_i}$ with cross-entropy loss. We find that infrastructure categories have high predictability between the infrastructure labels, show in Table \ref{baseline_results}. \emph{Electricity}, \emph{piped water}, and \emph{sewerage} achieve 0.89 AUROC.

These results offer a useful comparison for our model. If our model achieves performance similar to the oracle's, then its predictive power is as potent as if it predicted a set of concrete infrastructure labels that were correlated with the target outcome variable. Additionally, the oracle represents the predictive performance using \emph{expensive} and \emph{limited} survey data, whereas satellite imagery is \emph{cheap} and \emph{widely available}. Our best model on \emph{electricity}, \emph{piped water}, and \emph{road} performs comparably to the oracle. 

\section{Generalization Capabilities}

Ultimately, we are interested in deploying our deep learning approach to provide high-resolution maps of infrastructure quality that can be updated frequently based on relatively inexpensive remote sensing data. To this end, we evaluate the generalization capabilities of the model where we attempt to make predictions on data the model has not been explicitly trained on.

\subsection{Urban-Rural Split}
Each enumeration area in the Afrobarometer Round 6 survey is classified as being \emph{urban} or \emph{rural}. Urban and rural areas in Africa have significantly different infrastructural provisions. Urban areas are associated with improved water, and access to sanitation facilities is twice as great in urban areas compared with rural areas \cite{bentley2015inadequate}. Additionally, urban and rural areas have large visual differences in the satellite imagery that make them likely to be correlated with the other infrastructure metrics.

We measure the simple matching coefficient over all enumeration areas between the urban/rural variable and several infrastructure quality variables. Given binary variables $Y_i$ and $Y_j$, the simple matching coefficient of a sampled set of observations in $(Y_i, Y_j)$ is the proportion of matches in values between $Y_i$ and $Y_j$ divided by the number of samples in the set. The simple matching coefficient between \emph{urban/rural} and \emph{electricity}, \emph{sewerage}, and \emph{piped water} each is 0.70, 0.79, and 0.71 respectively. 

To address the concern that our model might be classifying the urban and rural indicators as a proxy for infrastructure quality, we evaluate the performance of our model on infrastructure variables within the \emph{urban} and \emph{rural} classes. Table \ref{afro_test_urban} shows the classification results on our best performing infrastructure metrics when the model is trained on only urban or rural areas. The AUROC scores are lower for all infrastructure variables, but not by enough to suggest that the model in the original classification task is exclusively learning to classify based on the urban/rural category. The AUROC of our highest performing outcome \emph{electricity} drops from 0.88 across both classes to 0.76 in the \emph{urban} class and 0.82 in the \emph{rural} class. 

\begin{table*}[!htb]
\centering
\caption{Results on \emph{electricity}, \emph{sewerage}, and \emph{piped water} when we stratify urban vs. rural areas. The model still performs well, indicating that it is not simply distinguishing urban and rural areas but is actually able to explain the variation within these classes.}
\label{afro_test_urban}
\begin{tabular}{|c|c|c|c|c|c|c|c|c|c|}
\hline
Satellite & Infrastructure & Urban/Rural & Balance & Accuracy & F1 Score & Precision & Recall & AUROC  \\
\hline 
L8 & Electricity & Urban & 0.953& 0.861 & 0.923& 0.971 & 0.880  & 0.763 \\
L8 & Electricity & Rural & 0.454& 0.741 & 0.724& 0.701 & 0.748 & 0.816 \\
L8 & Sewerage & Urban&0.661& 0.687 & 0.729& 0.853 & 0.636 & 0.794 \\
L8 & Sewerage & Rural&0.089& 0.897 & 0.430& 0.425 & 0.436 & 0.807 \\
L8 & Piped Water & Urban&0.861& 0.807 & 0.885& 0.907 & 0.864  & 0.758 \\
L8 & Piped Water & Rural&0.408& 0.628 & 0.599& 0.535 & 0.680  & 0.686 \\
\hline 
\end{tabular}
\end{table*}

\subsection{Country Hold-out}
In the original classification task, we sample our training and test sets uniformly among all 36 countries. With high probability, every country has data points that appear in the training set. However, we would also like to know whether deploying our model in an unobserved country leads to similarly strong classification results. 

We perform an experiment where we validate our model on new countries that it has not trained on before. In this experiment, we train on the enumeration areas of 35 countries, holding out one country, and then test our model on the enumeration areas of the held-out country. 

Table \ref{country_ho} shows the results on held-out countries Uganda, Tanzania, and Kenya, three of the countries with the most representation in the Afrobarometer survey. We train with strong regularization values to prevent overfitting on the trained countries. The results are not as strong as with the uniform sampling strategy; for example, we go from AUROC of 0.853 on \emph{electricity} on the test set when training with uniform sampling to AUROC of 0.637 on Ugandan enumeration areas when Uganda is held-out.

\begin{table}[h!]
\centering
\caption{Country hold-out results. We evaluate the performance of our model in a country not seen during training, simulating a realistic but challenging deployment situation. Compared to Table \ref{afro_test}, performance drops but the model maintains its usefulness for some infrastructure variables.}
\label{country_ho}
\begin{tabular}{cccccc}
\hline 
Country & Infrastructure & Balance & Accuracy & AUROC  \\
\hline
\hline 
Uganda & Electricity & 0.348 & 0.464 & 0.637 \\
 & Sewerage & 0.076 & 0.424  & 0.774 \\
 & Piped Water & 0.268 & 0.527 & 0.638 \\
Tanzania & Electricity & 0.521 & 0.500 & 0.541 \\
 & Sewerage & 0.103 & 0.502 & 0.578 \\
 & Piped Water & 0.432 & 0.445 & 0.588 \\
Kenya & Electricity & 0.846 & 0.703 & 0.518 \\
& Sewerage & 0.137 & 0.714 & 0.813 \\
& Piped Water & 0.418 & 0.473 & 0.602 \\
\hline
\end{tabular}
\end{table}
However, this is expected as the sample space of satellite images differ between countries, so enumeration areas between countries likely have geographic differences that make the salient features for classification less predictive.
\subsection{Fine-tuning Held-out Countries}
Though the results for the country hold-out experiments suggest that the model does not immediately generalize to new countries, we aim to show that transfer learning with a small, labeled sample in a new country generalizes those predictions to a significantly larger sample if the distribution of the training sample is representative. 

In this experiment, we repeat the procedure of the country hold-out experiments, but fine-tune the trained model on samples of the held-out countries. We train with L2 regularization of 0.1 and with different proportions of uniformly sampled data from the held-out country, from 0\% up to 80\%, where the lower end (0\%) is equivalent to the hold-out experiment, and the upper end (80\%) trains with the same amount of data as if the country was trained on in the initial training phase. We freeze the weight updates of the ResNet's parameters to obtain the final layer as visual features and then train a logistic regression with those features to predict the class label. We require the training and testing distribution to have the same proportion of positive labels.
\begin{figure}[h!]
\centering
\includegraphics[scale=0.60]{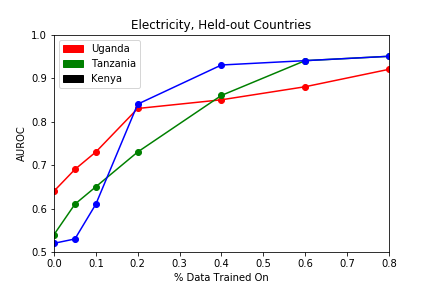}
\caption{AUROC scores on held-out countries when fine-tuned on samples of data. The x-axis corresponds to the percentage of data from the held-out country that the model was fine-tuned with.}
\label{electricity_holdout}
\end{figure}

Our best results on Uganda, Tanzania, and Kenya show that when each of these countries is held out, only 20\% of the country's data is needed to yield approximately the same AUROC as if the country was sampled uniformly in the training set. Additionally, training with 80\% of the country's data yields test scores as good as or better than the average of all the other countries that the model was trained on, with AUROC up to 0.96 and accuracy up to 92\%.

Figure \ref{electricity_holdout} shows the AUROC results for Uganda, Tanzania, and Kenya on the \emph{electricity} category as a function of the amount of data trained on. These results suggest that the model can be fine-tuned on limited data in a new country to good performance on a much larger test set in that country.

\section{Conclusion and Future Work}

Data on infrastructure quality outcomes in developing countries is lacking, and this work explored the use of globally available remote sensing data for predicting such outcomes. Using Afrobarometer survey data, we introduced a deep learning approach that demonstrates good predictive ability.

We experimented with Landsat 8 (multispectral) and Sentinel 1 (SAR) data, and obtained the best overall performance on Landsat 8. We believe the superior performance of Landsat 8 when compared to Sentinel 1 follows from Landsat 8 having RGB bands, allowing it to better use the ResNet's pretrained parameters. Sentinel 1 has no RGB bands.

We found the best performance on \emph{electricity}, \emph{sewerage}, \emph{piped water} and \emph{road}. Accuracy far surpasses balance and random prediction, and the AUROC scores are greater than 0.85 on \emph{electricity} and \emph{sewerage}. The model significantly outperforms the OSM and nearest neighbor interpolation baselines on these two variables by an average of 0.1 AUROC. Results also surpass the nightlights baseline. Furthermore, these results are on par with the oracle baseline, indicating that the model is making meaningful and accurate predictions. Intuitively, and as our models show, these variables are feature-defined structures and infrastructure systems that are uniquely distinguishable from satellite imagery. Figure \ref{best_vis} shows the distribution of all test set predictions for \emph{sewerage}, \emph{electricity}, and \emph{piped water}.

The first two images from left to right in Figure \ref{satellites} show sample satellite images where our model predicted a true positive and a false positive respectively for access to \emph{piped water}; the former shows clear indication of high activity with a large swath of developed buildings and roads, while the latter may have confused the model due to a high concentration of activity at the center of the image. Similarly the right two images show true and false positive predictions for \emph{electricity}. These predictions both demonstrate a similar proclivity for developed areas of buildings and roads.

We found poor performance on outcomes like \emph{market stalls}, \emph{health clinic}, and \emph{police station}; such outcomes barely outperformed random guessing and often underperformed OSM baseline results. This make sense as there are few features to resolve the presence of these particular buildings from satellite imagery, and OSM data may offer more insightful features.

The model exhibits high confidence in most predictions and training performance is significantly better than testing performance but we do not observe total overfitting. Hyperparameter tuning was not able to resolve these issues while maintaining optimal model performance. Turning towards the data, we found that images can appear highly similar even if they have different classifications. Possible solutions to this problem include both more data and deeper, more flexible models, although without sufficient data, the latter approach risks overfitting.

Our results demonstrate an exciting step forward in remote sensing and infrastructure mapping, far surpassing the OSM baseline. However, this task is presently underexplored, and we believe further improvements could be made. More local to the model, Sentinel 1 performance could likely be improved and more data could provide superior performance. Furthermore, transfer learning from other datasets, such as OSM, to these tasks offers a potential way to create a more effective model by learning to associate ground-level features and other observations with satellite imagery.  

Within the more general task of infrastructure mapping, we have also identified valuable future work. First, using the previous rounds of Afrobarometer, this model could be tested on its ability to generalize temporally. The ability to extrapolate how infrastructure has developed over time using contemporaneous imagery would be another exciting step in development. Second, a model that simultaneously trains using images from different satellites is worthy of further investigation. Third, the 10m and 30m resolution imagery used in this project is far from the resolution of today's satellites. We expect that higher resolution data would lead to better results and believe such an approach worthy of investigation.  Finally, a model that could take into account prior beliefs about infrastructure availability could offer a powerful tool for practical use.

For all these endeavors,  data will form a core issue. The quality of a deep model heavily relies on adequate data available, and a large focus should be towards making better use of existing image and survey data, through strong cataloging and collating efforts.  However, our results demonstrate the proof of concept that satellite imagery can be used to predict infrastructure quality.

\section{Acknowledgments}
We would like to acknowledge Zhongyi Tang, Hamza Husain, and George Azzari for support in data collection, and the Stanford Center on Global Poverty and Development for financial support. 



\bibliographystyle{ACM-Reference-Format}
\bibliography{egbib,liter_review_ref}


\begin{thebibliography}{43}


\ifx \showCODEN    \undefined \def \showCODEN     #1{\unskip}     \fi
\ifx \showDOI      \undefined \def \showDOI       #1{#1}\fi
\ifx \showISBNx    \undefined \def \showISBNx     #1{\unskip}     \fi
\ifx \showISBNxiii \undefined \def \showISBNxiii  #1{\unskip}     \fi
\ifx \showISSN     \undefined \def \showISSN      #1{\unskip}     \fi
\ifx \showLCCN     \undefined \def \showLCCN      #1{\unskip}     \fi
\ifx \shownote     \undefined \def \shownote      #1{#1}          \fi
\ifx \showarticletitle \undefined \def \showarticletitle #1{#1}   \fi
\ifx \showURL      \undefined \def \showURL       {\relax}        \fi
\providecommand\bibfield[2]{#2}
\providecommand\bibinfo[2]{#2}
\providecommand\natexlab[1]{#1}
\providecommand\showeprint[2][]{arXiv:#2}

\bibitem[\protect\citeauthoryear{Afrobarometer}{Afrobarometer}{2014}]%
        {AfrobarometerManual}
\bibfield{author}{\bibinfo{person}{Afrobarometer}.}
  \bibinfo{year}{2014}\natexlab{}.
\newblock \bibinfo{title}{Round 6 Survey Manual}.
\newblock   (\bibinfo{year}{2014}).
\newblock
\newblock
\shownote{http://www.afrobarometer.org/sites/default/files/survey\_\\manuals/ab\_r6\_survey\_manual\_en.pdf.}


\bibitem[\protect\citeauthoryear{Albert, Kaur, and Gonzalez}{Albert
  et~al\mbox{.}}{2017}]%
        {albert2017using}
\bibfield{author}{\bibinfo{person}{Adrian Albert}, \bibinfo{person}{Jasleen
  Kaur}, {and} \bibinfo{person}{Marta Gonzalez}.}
  \bibinfo{year}{2017}\natexlab{}.
\newblock \showarticletitle{Using convolutional networks and satellite imagery
  to identify patterns in urban environments at a large scale}.
\newblock \bibinfo{journal}{\emph{arXiv preprint arXiv:1704.02965}}
  (\bibinfo{year}{2017}).
\newblock


\bibitem[\protect\citeauthoryear{Audebert, Saux, and Lefèvre}{Audebert
  et~al\mbox{.}}{2017}]%
        {audebert_beyond_2017}
\bibfield{author}{\bibinfo{person}{Nicolas Audebert},
  \bibinfo{person}{Bertrand~Le Saux}, {and} \bibinfo{person}{Sébastien
  Lefèvre}.} \bibinfo{year}{2017}\natexlab{}.
\newblock \showarticletitle{Beyond RGB: Very high resolution urban remote
  sensing with multimodal deep networks}.
\newblock \bibinfo{journal}{\emph{ISPRS Journal of Photogrammetry and Remote
  Sensing}} (\bibinfo{year}{2017}).
\newblock
\showISSN{0924-2716}
\urldef\tempurl%
\url{https://doi.org/10.1016/j.isprsjprs.2017.11.011}
\showDOI{\tempurl}


\bibitem[\protect\citeauthoryear{Bentley, Han, and Houessou}{Bentley
  et~al\mbox{.}}{2015}]%
        {bentley2015inadequate}
\bibfield{author}{\bibinfo{person}{Thomas Bentley}, \bibinfo{person}{Kangwook
  Han}, {and} \bibinfo{person}{Richard Houessou}.}
  \bibinfo{year}{2015}\natexlab{}.
\newblock \showarticletitle{Inadequate access, poor government performance make
  water a top priority in Africa}.
\newblock  (\bibinfo{year}{2015}).
\newblock


\bibitem[\protect\citeauthoryear{BenYishay, Rotberg, Wells, Lv, Goodman,
  Kovacevic, and Runfola}{BenYishay et~al\mbox{.}}{2017}]%
        {benyishay2017geocoding}
\bibfield{author}{\bibinfo{person}{A BenYishay}, \bibinfo{person}{R Rotberg},
  \bibinfo{person}{J Wells}, \bibinfo{person}{Z Lv}, \bibinfo{person}{S
  Goodman}, \bibinfo{person}{L Kovacevic}, {and} \bibinfo{person}{D Runfola}.}
  \bibinfo{year}{2017}\natexlab{}.
\newblock \showarticletitle{Geocoding Afrobarometer rounds 1--6: methodology \&
  data quality}.
\newblock \bibinfo{journal}{\emph{AidData}} (\bibinfo{year}{2017}).
\newblock


\bibitem[\protect\citeauthoryear{Bragilevsky and Bajic}{Bragilevsky and
  Bajic}{[n. d.]}]%
        {bragilevskydeep}
\bibfield{author}{\bibinfo{person}{Lior Bragilevsky} {and}
  \bibinfo{person}{Ivan~V Bajic}.} \bibinfo{year}{[n. d.]}\natexlab{}.
\newblock \showarticletitle{Deep Learning for Amazon Satellite Image Analysis}.
\newblock  (\bibinfo{year}{[n. d.]}).
\newblock


\bibitem[\protect\citeauthoryear{Castelluccio, Poggi, Sansone, and
  Verdoliva}{Castelluccio et~al\mbox{.}}{2015}]%
        {Castelluccio2015land}
\bibfield{author}{\bibinfo{person}{Marco Castelluccio},
  \bibinfo{person}{Giovanni Poggi}, \bibinfo{person}{Carlo Sansone}, {and}
  \bibinfo{person}{Luisa Verdoliva}.} \bibinfo{year}{2015}\natexlab{}.
\newblock \showarticletitle{Land use classification in remote sensing images by
  convolutional neural networks}.
\newblock \bibinfo{journal}{\emph{arXiv preprint arXiv:1508.00092}}
  (\bibinfo{year}{2015}).
\newblock


\bibitem[\protect\citeauthoryear{Center}{Center}{2013}]%
        {dmsp}
\bibfield{author}{\bibinfo{person}{National Geophysical~Data Center}.}
  \bibinfo{year}{2013}\natexlab{}.
\newblock \bibinfo{title}{Version 4 DMSP-OLS Nighttime Lights Time Series}.
\newblock   (\bibinfo{year}{2013}).
\newblock
\newblock
\shownote{https://ngdc.noaa.gov/eog/dmsp/downloadV4composites.html.}


\bibitem[\protect\citeauthoryear{Center}{Center}{2015}]%
        {viirs}
\bibfield{author}{\bibinfo{person}{National Geophysical~Data Center}.}
  \bibinfo{year}{2015}\natexlab{}.
\newblock \bibinfo{title}{Version 1 VIIRS Day/Night Band Nighttime Lights}.
\newblock   (\bibinfo{year}{2015}).
\newblock
\newblock
\shownote{https://ngdc.noaa.gov/eog/viirs/index.html.}


\bibitem[\protect\citeauthoryear{Christopher~Brown}{Christopher~Brown}{[n.
  d.]}]%
        {ROC}
\bibfield{author}{\bibinfo{person}{Herbert~Davis Christopher~Brown}.}
  \bibinfo{year}{[n. d.]}\natexlab{}.
\newblock \bibinfo{title}{Receiver operating characteristics curves and related
  decision measures: A tutorial}.
\newblock   (\bibinfo{year}{[n. d.]}).
\newblock


\bibitem[\protect\citeauthoryear{Dabalen, Etang, Hoogeveen, Mushi, Schipper,
  and von Engelhardt}{Dabalen et~al\mbox{.}}{2016}]%
        {dabalen2016mobile}
\bibfield{author}{\bibinfo{person}{Andrew Dabalen}, \bibinfo{person}{Alvin
  Etang}, \bibinfo{person}{Johannes Hoogeveen}, \bibinfo{person}{Elvis Mushi},
  \bibinfo{person}{Youdi Schipper}, {and} \bibinfo{person}{Johannes von
  Engelhardt}.} \bibinfo{year}{2016}\natexlab{}.
\newblock \bibinfo{booktitle}{\emph{Mobile Phone Panel Surveys in Developing
  Countries: A Practical Guide for Microdata Collection}}.
\newblock \bibinfo{publisher}{World Bank Publications}.
\newblock


\bibitem[\protect\citeauthoryear{Dai, Li, He, and Sun}{Dai
  et~al\mbox{.}}{2016}]%
        {dai2016r}
\bibfield{author}{\bibinfo{person}{Jifeng Dai}, \bibinfo{person}{Yi Li},
  \bibinfo{person}{Kaiming He}, {and} \bibinfo{person}{Jian Sun}.}
  \bibinfo{year}{2016}\natexlab{}.
\newblock \showarticletitle{R-fcn: Object detection via region-based fully
  convolutional networks}. In \bibinfo{booktitle}{\emph{Advances in neural
  information processing systems}}. \bibinfo{pages}{379--387}.
\newblock


\bibitem[\protect\citeauthoryear{Esteva, Kuprel, Novoa, Ko, Swetter, Blau, and
  Thrun}{Esteva et~al\mbox{.}}{2017}]%
        {esteva2017dermatologist}
\bibfield{author}{\bibinfo{person}{Andre Esteva}, \bibinfo{person}{Brett
  Kuprel}, \bibinfo{person}{Roberto~A Novoa}, \bibinfo{person}{Justin Ko},
  \bibinfo{person}{Susan~M Swetter}, \bibinfo{person}{Helen~M Blau}, {and}
  \bibinfo{person}{Sebastian Thrun}.} \bibinfo{year}{2017}\natexlab{}.
\newblock \showarticletitle{Dermatologist-level classification of skin cancer
  with deep neural networks}.
\newblock \bibinfo{journal}{\emph{Nature}} \bibinfo{volume}{542},
  \bibinfo{number}{7639} (\bibinfo{year}{2017}), \bibinfo{pages}{115}.
\newblock


\bibitem[\protect\citeauthoryear{Glorot and Bengio}{Glorot and Bengio}{2010}]%
        {glorot2010understanding}
\bibfield{author}{\bibinfo{person}{Xavier Glorot} {and} \bibinfo{person}{Yoshua
  Bengio}.} \bibinfo{year}{2010}\natexlab{}.
\newblock \showarticletitle{Understanding the difficulty of training deep
  feedforward neural networks}. In \bibinfo{booktitle}{\emph{Proceedings of the
  Thirteenth International Conference on Artificial Intelligence and
  Statistics}}. \bibinfo{pages}{249--256}.
\newblock


\bibitem[\protect\citeauthoryear{Haklay and Weber}{Haklay and Weber}{2008}]%
        {haklay2008openstreetmap}
\bibfield{author}{\bibinfo{person}{Mordechai Haklay} {and}
  \bibinfo{person}{Patrick Weber}.} \bibinfo{year}{2008}\natexlab{}.
\newblock \showarticletitle{Openstreetmap: User-generated street maps}.
\newblock \bibinfo{journal}{\emph{IEEE Pervasive Computing}}
  \bibinfo{volume}{7}, \bibinfo{number}{4} (\bibinfo{year}{2008}),
  \bibinfo{pages}{12--18}.
\newblock


\bibitem[\protect\citeauthoryear{He, Zhang, Ren, and Sun}{He
  et~al\mbox{.}}{2016}]%
        {he2016deep}
\bibfield{author}{\bibinfo{person}{Kaiming He}, \bibinfo{person}{Xiangyu
  Zhang}, \bibinfo{person}{Shaoqing Ren}, {and} \bibinfo{person}{Jian Sun}.}
  \bibinfo{year}{2016}\natexlab{}.
\newblock \showarticletitle{Deep residual learning for image recognition}. In
  \bibinfo{booktitle}{\emph{Proceedings of the IEEE conference on computer
  vision and pattern recognition}}. \bibinfo{pages}{770--778}.
\newblock


\bibitem[\protect\citeauthoryear{Helbich, Amelunxen, Neis, and Zipf}{Helbich
  et~al\mbox{.}}{2012}]%
        {helbich2012comparative}
\bibfield{author}{\bibinfo{person}{Marco Helbich}, \bibinfo{person}{Chritoph
  Amelunxen}, \bibinfo{person}{Pascal Neis}, {and} \bibinfo{person}{Alexander
  Zipf}.} \bibinfo{year}{2012}\natexlab{}.
\newblock \showarticletitle{Comparative spatial analysis of positional accuracy
  of OpenStreetMap and proprietary geodata}.
\newblock \bibinfo{journal}{\emph{Proceedings of GI\_Forum}}
  (\bibinfo{year}{2012}), \bibinfo{pages}{24--33}.
\newblock


\bibitem[\protect\citeauthoryear{IEAG}{IEAG}{2014}]%
        {ieag2014world}
\bibfield{author}{\bibinfo{person}{UN IEAG}.} \bibinfo{year}{2014}\natexlab{}.
\newblock \bibinfo{title}{A World that Counts--Mobilising the Data Revolution
  for Sustainable Development}.
\newblock   (\bibinfo{year}{2014}).
\newblock


\bibitem[\protect\citeauthoryear{Irish}{Irish}{2000}]%
        {irish2000landsat}
\bibfield{author}{\bibinfo{person}{Richard~R Irish}.}
  \bibinfo{year}{2000}\natexlab{}.
\newblock \showarticletitle{Landsat 7 automatic cloud cover assessment}. In
  \bibinfo{booktitle}{\emph{Algorithms for Multispectral, Hyperspectral, and
  Ultraspectral Imagery VI}}, Vol.~\bibinfo{volume}{4049}. International
  Society for Optics and Photonics, \bibinfo{pages}{348--356}.
\newblock


\bibitem[\protect\citeauthoryear{Jean, Burke, Xie, Davis, Lobell, and
  Ermon}{Jean et~al\mbox{.}}{2016}]%
        {jean2016combining}
\bibfield{author}{\bibinfo{person}{Neal Jean}, \bibinfo{person}{Marshall
  Burke}, \bibinfo{person}{Michael Xie}, \bibinfo{person}{W.~Matthew Davis},
  \bibinfo{person}{David Lobell}, {and} \bibinfo{person}{Stefano Ermon}.}
  \bibinfo{year}{2016}\natexlab{}.
\newblock \showarticletitle{Combining satellite imagery and machine learning to
  predict poverty}.
\newblock \bibinfo{journal}{\emph{Science}} \bibinfo{volume}{353},
  \bibinfo{number}{3601} (\bibinfo{year}{2016}), \bibinfo{pages}{790--794}.
\newblock


\bibitem[\protect\citeauthoryear{Jerven}{Jerven}{2014}]%
        {jerven2014benefits}
\bibfield{author}{\bibinfo{person}{Morten Jerven}.}
  \bibinfo{year}{2014}\natexlab{}.
\newblock \showarticletitle{Benefits and costs of the data for development
  targets for the Post-2015 Development Agenda}.
\newblock \bibinfo{journal}{\emph{Data for Development Assessment Paper Working
  Paper, September. Copenhagen: Copenhagen Consensus Center}}
  (\bibinfo{year}{2014}).
\newblock


\bibitem[\protect\citeauthoryear{Kingma and Ba}{Kingma and Ba}{2014}]%
        {kingma2014adam}
\bibfield{author}{\bibinfo{person}{Diederik~P Kingma} {and}
  \bibinfo{person}{Jimmy Ba}.} \bibinfo{year}{2014}\natexlab{}.
\newblock \showarticletitle{Adam: A method for stochastic optimization}.
\newblock \bibinfo{journal}{\emph{arXiv preprint arXiv:1412.6980}}
  (\bibinfo{year}{2014}).
\newblock


\bibitem[\protect\citeauthoryear{Krizhevsky, Sutskever, and Hinton}{Krizhevsky
  et~al\mbox{.}}{2012}]%
        {krizhevsky2012imagenet}
\bibfield{author}{\bibinfo{person}{Alex Krizhevsky}, \bibinfo{person}{Ilya
  Sutskever}, {and} \bibinfo{person}{Geoffrey~E Hinton}.}
  \bibinfo{year}{2012}\natexlab{}.
\newblock \showarticletitle{Imagenet classification with deep convolutional
  neural networks}. In \bibinfo{booktitle}{\emph{Advances in neural information
  processing systems}}. \bibinfo{pages}{1097--1105}.
\newblock


\bibitem[\protect\citeauthoryear{Liu, Hang, Song, and Li}{Liu
  et~al\mbox{.}}{2017}]%
        {liu2017learning}
\bibfield{author}{\bibinfo{person}{Qingshan Liu}, \bibinfo{person}{Renlong
  Hang}, \bibinfo{person}{Huihui Song}, {and} \bibinfo{person}{Zhi Li}.}
  \bibinfo{year}{2017}\natexlab{}.
\newblock \showarticletitle{Learning Multiscale Deep Features for
  High-Resolution Satellite Image Scene Classification}.
\newblock \bibinfo{journal}{\emph{IEEE Transactions on Geoscience and Remote
  Sensing}} (\bibinfo{year}{2017}).
\newblock


\bibitem[\protect\citeauthoryear{Maharana, Nguyen, and Nsoesie}{Maharana
  et~al\mbox{.}}{2017}]%
        {DBLP:journals/corr/abs-1710-05483}
\bibfield{author}{\bibinfo{person}{Adyasha Maharana}, \bibinfo{person}{Quynh~C.
  Nguyen}, {and} \bibinfo{person}{Elaine~O. Nsoesie}.}
  \bibinfo{year}{2017}\natexlab{}.
\newblock \showarticletitle{Using Deep Learning and Satellite Imagery to
  Quantify the Impact of the Built Environment on Neighborhood Crime Rates}.
\newblock \bibinfo{journal}{\emph{CoRR}}  \bibinfo{volume}{abs/1710.05483}
  (\bibinfo{year}{2017}).
\newblock
\showeprint[arxiv]{1710.05483}
\urldef\tempurl%
\url{http://arxiv.org/abs/1710.05483}
\showURL{%
\tempurl}


\bibitem[\protect\citeauthoryear{Mnih and Hinton}{Mnih and Hinton}{2010}]%
        {mnih2010learning}
\bibfield{author}{\bibinfo{person}{Volodymyr Mnih} {and}
  \bibinfo{person}{Geoffrey~E Hinton}.} \bibinfo{year}{2010}\natexlab{}.
\newblock \showarticletitle{Learning to detect roads in high-resolution aerial
  images}. In \bibinfo{booktitle}{\emph{European Conference on Computer
  Vision}}. Springer, \bibinfo{pages}{210--223}.
\newblock


\bibitem[\protect\citeauthoryear{Mnih and Hinton}{Mnih and Hinton}{2012}]%
        {mnih2012learning}
\bibfield{author}{\bibinfo{person}{Volodymyr Mnih} {and}
  \bibinfo{person}{Geoffrey~E Hinton}.} \bibinfo{year}{2012}\natexlab{}.
\newblock \showarticletitle{Learning to label aerial images from noisy data}.
  In \bibinfo{booktitle}{\emph{Proceedings of the 29th International Conference
  on Machine Learning (ICML-12)}}. \bibinfo{pages}{567--574}.
\newblock


\bibitem[\protect\citeauthoryear{OECD}{OECD}{2014}]%
        {/content/publication/9789264217294-en}
\bibfield{author}{\bibinfo{person}{OECD}.} \bibinfo{year}{2014}\natexlab{}.
\newblock \bibinfo{booktitle}{\emph{The Space Economy at a Glance 2014}}.
\newblock 144 pages.
\newblock
\urldef\tempurl%
\url{https://doi.org/http://dx.doi.org/10.1787/9789264217294-en}
\showDOI{\tempurl}


\bibitem[\protect\citeauthoryear{Oquab, Bottou, Laptev, and Sivic}{Oquab
  et~al\mbox{.}}{2014}]%
        {oquab2014learning}
\bibfield{author}{\bibinfo{person}{Maxime Oquab}, \bibinfo{person}{Leon
  Bottou}, \bibinfo{person}{Ivan Laptev}, {and} \bibinfo{person}{Josef Sivic}.}
  \bibinfo{year}{2014}\natexlab{}.
\newblock \showarticletitle{Learning and transferring mid-level image
  representations using convolutional neural networks}. In
  \bibinfo{booktitle}{\emph{Computer Vision and Pattern Recognition (CVPR),
  2014 IEEE Conference on}}. IEEE, \bibinfo{pages}{1717--1724}.
\newblock


\bibitem[\protect\citeauthoryear{Oyuke, Penar, and Howard}{Oyuke
  et~al\mbox{.}}{[n. d.]}]%
        {AfrobarometerElectricityReport}
\bibfield{author}{\bibinfo{person}{Abel Oyuke}, \bibinfo{person}{Peter~Halley
  Penar}, {and} \bibinfo{person}{Brian Howard}.} \bibinfo{year}{[n.
  d.]}\natexlab{}.
\newblock \bibinfo{title}{Afrobarometer Dispatch No.75}.
\newblock   (\bibinfo{year}{[n. d.]}).
\newblock


\bibitem[\protect\citeauthoryear{Papadomanolaki, Vakalopoulou, Zagoruyko, and
  Karantzalos}{Papadomanolaki et~al\mbox{.}}{2016}]%
        {papadomanolaki2016benchmarking}
\bibfield{author}{\bibinfo{person}{M Papadomanolaki}, \bibinfo{person}{M
  Vakalopoulou}, \bibinfo{person}{S Zagoruyko}, {and} \bibinfo{person}{K
  Karantzalos}.} \bibinfo{year}{2016}\natexlab{}.
\newblock \showarticletitle{BENCHMARKING DEEP LEARNING FRAMEWORKS FOR THE
  CLASSIFICATION OF VERY HIGH RESOLUTION SATELLITE MULTISPECTRAL DATA.}
\newblock \bibinfo{journal}{\emph{ISPRS Annals of Photogrammetry, Remote
  Sensing \& Spatial Information Sciences}} \bibinfo{volume}{3},
  \bibinfo{number}{7} (\bibinfo{year}{2016}).
\newblock


\bibitem[\protect\citeauthoryear{Penatti, Nogueira, and dos Santos}{Penatti
  et~al\mbox{.}}{2015}]%
        {penatti2015deep}
\bibfield{author}{\bibinfo{person}{Ot{\'a}vio~AB Penatti},
  \bibinfo{person}{Keiller Nogueira}, {and} \bibinfo{person}{Jefersson~A dos
  Santos}.} \bibinfo{year}{2015}\natexlab{}.
\newblock \showarticletitle{Do deep features generalize from everyday objects
  to remote sensing and aerial scenes domains?}. In
  \bibinfo{booktitle}{\emph{Proceedings of the IEEE Conference on Computer
  Vision and Pattern Recognition Workshops}}. \bibinfo{pages}{44--51}.
\newblock


\bibitem[\protect\citeauthoryear{Pottas}{Pottas}{2014}]%
        {pottas2014addressing}
\bibfield{author}{\bibinfo{person}{Andr{\'e} Pottas}.}
  \bibinfo{year}{2014}\natexlab{}.
\newblock \showarticletitle{Addressing Africa’s Infrastructure Challenges}.
\newblock \bibinfo{journal}{\emph{Delloitte. Accessed online on}}
  (\bibinfo{year}{2014}).
\newblock


\bibitem[\protect\citeauthoryear{Pryzant, Ermon, and Lobell}{Pryzant
  et~al\mbox{.}}{2017}]%
        {pryzant2017monitoring}
\bibfield{author}{\bibinfo{person}{Reid Pryzant}, \bibinfo{person}{Stefano
  Ermon}, {and} \bibinfo{person}{David Lobell}.}
  \bibinfo{year}{2017}\natexlab{}.
\newblock \showarticletitle{Monitoring Ethiopian Wheat Fungus with Satellite
  Imagery and Deep Feature Learning}. In \bibinfo{booktitle}{\emph{2017 IEEE
  Conference on Computer Vision and Pattern Recognition Workshops (CVPRW)}}.
  IEEE, \bibinfo{pages}{1524--1532}.
\newblock


\bibitem[\protect\citeauthoryear{Reibel}{Reibel}{2007}]%
        {reibel2007geographic}
\bibfield{author}{\bibinfo{person}{Michael Reibel}.}
  \bibinfo{year}{2007}\natexlab{}.
\newblock \showarticletitle{Geographic information systems and spatial data
  processing in demography: a review}.
\newblock \bibinfo{journal}{\emph{Population Research and Policy Review}}
  \bibinfo{volume}{26}, \bibinfo{number}{5-6} (\bibinfo{year}{2007}),
  \bibinfo{pages}{601--618}.
\newblock


\bibitem[\protect\citeauthoryear{Romero, Gatta, and Camps-Valls}{Romero
  et~al\mbox{.}}{2016}]%
        {romero2016unsupervised}
\bibfield{author}{\bibinfo{person}{Adriana Romero}, \bibinfo{person}{Carlo
  Gatta}, {and} \bibinfo{person}{Gustau Camps-Valls}.}
  \bibinfo{year}{2016}\natexlab{}.
\newblock \showarticletitle{Unsupervised deep feature extraction for remote
  sensing image classification}.
\newblock \bibinfo{journal}{\emph{IEEE Transactions on Geoscience and Remote
  Sensing}} \bibinfo{volume}{54}, \bibinfo{number}{3} (\bibinfo{year}{2016}),
  \bibinfo{pages}{1349--1362}.
\newblock


\bibitem[\protect\citeauthoryear{Sandefur and Glassman}{Sandefur and
  Glassman}{2015}]%
        {sandefur2015political}
\bibfield{author}{\bibinfo{person}{Justin Sandefur} {and}
  \bibinfo{person}{Amanda Glassman}.} \bibinfo{year}{2015}\natexlab{}.
\newblock \showarticletitle{The political economy of bad data: evidence from
  African survey and administrative statistics}.
\newblock \bibinfo{journal}{\emph{The Journal of Development Studies}}
  \bibinfo{volume}{51}, \bibinfo{number}{2} (\bibinfo{year}{2015}),
  \bibinfo{pages}{116--132}.
\newblock


\bibitem[\protect\citeauthoryear{Simonyan and Zisserman}{Simonyan and
  Zisserman}{2014}]%
        {simonyan2014very}
\bibfield{author}{\bibinfo{person}{Karen Simonyan} {and}
  \bibinfo{person}{Andrew Zisserman}.} \bibinfo{year}{2014}\natexlab{}.
\newblock \showarticletitle{Very deep convolutional networks for large-scale
  image recognition}.
\newblock \bibinfo{journal}{\emph{arXiv preprint arXiv:1409.1556}}
  (\bibinfo{year}{2014}).
\newblock


\bibitem[\protect\citeauthoryear{Varshney, Chen, Abelson, Nowocin, Sakhrani,
  Xu, and Spatocco}{Varshney et~al\mbox{.}}{2015}]%
        {varshney2015targeting}
\bibfield{author}{\bibinfo{person}{Kush~R Varshney}, \bibinfo{person}{George~H
  Chen}, \bibinfo{person}{Brian Abelson}, \bibinfo{person}{Kendall Nowocin},
  \bibinfo{person}{Vivek Sakhrani}, \bibinfo{person}{Ling Xu}, {and}
  \bibinfo{person}{Brian~L Spatocco}.} \bibinfo{year}{2015}\natexlab{}.
\newblock \showarticletitle{Targeting villages for rural development using
  satellite image analysis}.
\newblock \bibinfo{journal}{\emph{Big Data}} \bibinfo{volume}{3},
  \bibinfo{number}{1} (\bibinfo{year}{2015}), \bibinfo{pages}{41--53}.
\newblock


\bibitem[\protect\citeauthoryear{Xie, Jean, Burke, Lobell, and Ermon}{Xie
  et~al\mbox{.}}{2016}]%
        {xie2016transfer}
\bibfield{author}{\bibinfo{person}{Michael Xie}, \bibinfo{person}{Neal Jean},
  \bibinfo{person}{Marshall Burke}, \bibinfo{person}{David Lobell}, {and}
  \bibinfo{person}{Stefano Ermon}.} \bibinfo{year}{2016}\natexlab{}.
\newblock \showarticletitle{Transfer Learning from Deep Features for Remote
  Sensing and Poverty Mapping}.
\newblock \bibinfo{journal}{\emph{arXiv preprint arXiv:1510.00098}}
  (\bibinfo{year}{2016}).
\newblock


\bibitem[\protect\citeauthoryear{Yang and Newsam}{Yang and Newsam}{2010}]%
        {yang2010bag}
\bibfield{author}{\bibinfo{person}{Yi Yang} {and} \bibinfo{person}{Shawn
  Newsam}.} \bibinfo{year}{2010}\natexlab{}.
\newblock \showarticletitle{Bag-of-visual-words and spatial extensions for
  land-use classification}. In \bibinfo{booktitle}{\emph{Proceedings of the
  18th SIGSPATIAL international conference on advances in geographic
  information systems}}. ACM, \bibinfo{pages}{270--279}.
\newblock


\bibitem[\protect\citeauthoryear{You, Li, Low, Lobell, and Ermon}{You
  et~al\mbox{.}}{2017}]%
        {you2017deep}
\bibfield{author}{\bibinfo{person}{Jiaxuan You}, \bibinfo{person}{Xiaocheng
  Li}, \bibinfo{person}{Melvin Low}, \bibinfo{person}{David Lobell}, {and}
  \bibinfo{person}{Stefano Ermon}.} \bibinfo{year}{2017}\natexlab{}.
\newblock \showarticletitle{Deep Gaussian Process for Crop Yield Prediction
  Based on Remote Sensing Data.}. In \bibinfo{booktitle}{\emph{AAAI}}.
  \bibinfo{pages}{4559--4566}.
\newblock


\bibitem[\protect\citeauthoryear{Yuan}{Yuan}{2016}]%
        {yuan2016automatic}
\bibfield{author}{\bibinfo{person}{Jiangye Yuan}.}
  \bibinfo{year}{2016}\natexlab{}.
\newblock \showarticletitle{Automatic building extraction in aerial scenes
  using convolutional networks}.
\newblock \bibinfo{journal}{\emph{arXiv preprint arXiv:1602.06564}}
  (\bibinfo{year}{2016}).
\newblock


\end{thebibliography}

\end{document}